\documentclass[fleqn,usenatbib]{mnras}

\usepackage{newtxtext,newtxmath}

\usepackage[T1]{fontenc}

\DeclareRobustCommand{\VAN}[3]{#2}
\let\VANthebibliography\thebibliography
\def\thebibliography{\DeclareRobustCommand{\VAN}[3]{##3}\VANthebibliography}

\usepackage{graphicx}	
\usepackage{amsmath}	
\usepackage{ulem}
\usepackage{comment}
\usepackage{subcaption}

\newcommand{\gwcosmo}{\texttt{gwcosmo }}

\title[Dark sirens with bright subsets]{Dark standard siren cosmology with bright galaxy subsets}

\author[K. Naveed et al.]{
Khuzaifa Naveed,$^{1}$\thanks{E-mail: muhammadkhuzaifa.naveed@ugent.be}
Cezary Turski,$^{1}$
and Archisman Ghosh$^{1}$
\\
$^{1}$Department of Physics \& Astronomy, Ghent University, Proeftuinstraat 86, 9000 Ghent, Belgium\\
}

\date{Accepted XXX. Received YYY; in original form ZZZ}

\pubyear{\the\year{}}

\begin{document}
\label{firstpage}
\pagerange{\pageref{firstpage}--\pageref{lastpage}}
\maketitle

\begin{abstract}
In this short paper, we investigate the impact of selecting only a subset of bright galaxies to provide redshift information for a dark standard siren measurement of the Hubble constant $H_0$. Employing gravitational-wave observations from the Third Gravitational-Wave Transient Catalogue (GWTC-3) in conjunction with the GLADE+ galaxy catalogue, we show that restricting to bright galaxy subsets can enhance the precision of the $H_0$ estimate by up to $80\%$ in the most favourable scenario. A comprehensive assessment of systematic uncertainties is still required. This work lays the foundation for employing alternative tracers --- such as brightest cluster galaxies (BCGs) and luminous red galaxies (LRGs) --- in gravitational-wave cosmology, particularly at redshifts where conventional galaxy catalogues offer limited coverage.

\end{abstract}




\section{Introduction}

A disturbing puzzle in current observational cosmology is the ongoing tension between measurements of the Hubble constant $H_0$, obtained from observations of the early universe (e.g. Planck CMB, \cite{Planck:2018vyg}) and those from the local universe (e.g. Cepheids and Type Ia supernovae, \cite{Riess:2019cxk}). This discrepancy, exceeding the $5\sigma$ threshold, has sparked debate over whether it reflects systematic errors or reveals physics beyond the standard cosmological model. 

Gravitational waves (GWs) from compact binary coalescences offer a unique, independent way to measure $H_0$. These sources act as ``standard sirens'' with the waveform directly encoding the luminosity distance, circumventing traditional cosmic distance ladders~\citep{Schutz:1986gp,Holz:2005df}. When a GW event is accompanied by an electromagnetic (EM) counterpart, as for the binary neutron star merger GW170817, the redshift of the host galaxy can be directly measured, enabling a direct $H_0$ measurement~\citep{LIGOScientific:2017adf}. However, such bright siren events are rare. 

In contrast, the vast majority of GW detections to date are ``dark sirens,'' lacking confirmed electromagnetic (EM) counterparts. For these events, redshift information must be inferred statistically by cross-matching sky localizations with galaxy catalogues.\footnote{In a parallel ``spectral siren'' approach, the redshift information comes from the observed mass distribution of merging compact binaries~\citep{Farr:2019twy,Mastrogiovanni:2021wsd,Ezquiaga:2022zkx,Agarwal:2024hld}. The same mass distribution enters the selection function for the dark siren analysis, and the unknown parameters of the distribution need to be carefully marginalized over~\citep{Mastrogiovanni:2023emh,Mastrogiovanni:2023zbw,Gray:2023wgj}. A comprehensive review of the subject, including other approaches, can be found in \cite{Chen:2024gdn,Palmese:2025zku}. Some common misconceptions about the dark standard siren method have been addressed in \cite{Gair:2022zsa}.} This method, originally proposed by \cite{Schutz:1986gp} and further developed and applied to LIGO-Virgo-KAGRA (LVK) data in \cite{LIGOScientific:2018gmd,DES:2019ccw,LIGOScientific:2019zcs,LIGOScientific:2021aug,Finke:2021aom,Palmese:2021mjm,Bom:2024afj}, enables cosmological inference using all detectable mergers, vastly increasing the effective event sample. 

However, a major limitation in current dark siren analyses arises from the incompleteness of galaxy catalogues, especially at higher redshifts. This hampers precision in the $H_0$ measurement and makes it more susceptible to biases coming from the construction of the out-of-catalogue part, namely, the modelled contribution of the faint galaxies missing from the catalogue. In this work, we introduce a refined dark siren approach that addresses this challenge by using only the brightest galaxies as redshift tracers. Bright galaxies are easier to detect and visible out to higher redshifts, allowing us to reduce the contribution of uncertain, faint out-of-catalogue galaxies, potentially improving both accuracy and precision in cosmological inference.

The idea behind choosing only bright galaxies is essentially that brightest cluster galaxies (BCGs) of galaxy clusters as well as luminous red galaxies (LRGs) trace the underlying matter distribution --- the nodes and the filaments of the cosmic web. While existing galaxy catalogues are severely incomplete at redshifts beyond $z\gtrsim0.5$, these bright objects can be observed to much farther away ($z\gtrsim6$), mapping crudely the 3D locations of the potential GW events in redshift space. Given large uncertain uncertainties in GW distance estimates ($\mathcal{O}(10\%)$), the small error incurred by replacing the redshift of a host galaxy with that a nearby bright galaxy, is not expected to have a significant drawback. In this sense, the use of such sporadic bright tracers extends the concept of highly-informative ``golden'' dark sirens developed in \cite{Borhanian:2020vyr,Gupta:2022fwd} to a wider class of GW events. Moreover, the idea of luminous objects tracing the underlying matter distribution also lies at the core of cross-correlation techniques for cosmography, see e.g., \cite{Diaz:2021pem,Mukherjee:2022afz,Afroz:2024joi,Ghosh:2023ksl}.

In this work, we use the \gwcosmo pipeline~\citep{Gray:2019ksv,Gray:2021sew,Gray:2023wgj} for dark siren analysis. Each galaxy in the catalogue is assigned a weight reflecting its likelihood of hosting a merger event, based on its inferred stellar mass and redshift. These host probability weights are typically proportional to luminosity in a chosen band, commonly the near-infrared $K$-band, as a proxy for stellar mass. This enables construction of a redshift prior along each line of sight by summing contributions from galaxies in the catalogue (the ``in-catalogue'' part), and augmenting with an analytic model to capture fainter galaxies not in the galaxy catalogue (the ``out-of-catalogue'' part).

Our approach builds on this framework by selecting only the top 
$XX\%$ of the brightest galaxies (ranked by their absolute luminosity obtained in the $K$-band), under the simplifying assumption that these galaxies trace the large-scale matter distribution. By focusing on the most luminous (and thus more complete) portion of the catalogue, we reduce the contribution of uncertain, faint galaxies while preserving sensitivity to the cosmic density field. This potentially enables a more robust cosmological inference, maximizing the utility of current galaxy catalogues without introducing additional dependencies.

We apply this technique to the full set of confident detections from GWTC‑3~\citep{KAGRA:2021vkt}, constructing line-of-sight redshift priors that blend in-catalogue and out-of-catalogue parts. By progressively selecting the top $XX\%$ of brightest galaxies in the GLADE+ catalogue~\citep{Dalya:2021ewn}, we examine how catalogue completeness and depth trade off against systematic uncertainties. 
Our results demonstrate that with current data, bright subsets can yield improved constraints on $H_0$, paving the way for more robust dark siren cosmology with distant GW events and alternative EM tracers.

While this manuscript was undergoing peer-review, a related work, \cite{VanWyngarden:2025ogy} appeared. It considers the mock galaxy catalogue MICECAT \citep{Fosalba:2013wxa} and simplified GW simulations (with Gaussian uncertainties) to demonstrate that with as low as $1\%$ of the brightest galaxies in the catalogue, one can recover an unbiased estimate in $H_0$ with such datasets. Our work is complementary and explores similar effects on real data.

The rest of this paper is organized as follows. We begin with the GW and EM data used in this work in Section~\ref{sec:data} and give a brief overview of our method in Section~\ref{sec:method}. In Section~\ref{sec:angular_power}, we demonstrate that subselecting the brightest galaxies does not significantly alter the underlying large-scale structure probed by these tracers. Our results on GWTC-3 data are presented in Section~\ref{sec:results}, where we also discuss robustness studies involving the effect of luminosity weighting.
We conclude in Section~\ref{sec:conclusions} with an outlook towards the future.

\section{Data}
\label{sec:data}
\subsection{GW data}
\label{subsec:gw_data}
We analyze GW events published by the LVK collaboration in GWTC-3 GW transient catalogue.\footnote{Posterior samples and skymaps for events observed up to the first half of the third observing run (O3a) are available for download at \url{https://zenodo.org/records/6513631}. Data corresponding to the latter half of the run (O3b) can be found at \url{https://zenodo.org/records/5546663}.} We apply the cut in the signal-to-noise ratio $\text{SNR}>11$, which leaves us with 47 events: 42 BBHs, 3 NSBHs, and 2 BNSs. The same set of events have been analyzed as in \cite{LIGOScientific:2021aug}; however we leave out the bright standard siren GW170817. In this paper we assume that compact binary coalescences (CBCs) follow a Madau-Dickinson merger redshift evolution model~\citep{Madau:2014bja} with the high-$z$ slope $k=2.86$, low-$z$ slope $\gamma=4.59$ and the break-point $z_p=2.47$. For BBHs we use power-law with a Gaussian peak mass model, with the power-law slope $\alpha=3.78$, the mean of the Gaussian peak $\sigma_g=3.88$, with width $\mu_g=32.27$, and the relative weight between power-law and the peak $\lambda_g=0.03$. We assume that the mass of a primary BH is between $M_{\text{min}}=4.98\,M_{\odot}$ and $M_{\text{max}}=112.5\,M_{\odot}$. We assume that mass of NSs is uniformly distributed between $M_{\text{min, NS}}= 1.0\,M_{\odot}$ and $M_{\text{max, NS}}=3.0\,M_{\odot}$. These fixed parameters can easily be varied or marginalized over in a more realistic study~\citep{Gray:2023wgj}.

\subsection{EM data}

To provide statistical redshift information (“dark siren” method) we employ the all‑sky GLADE+ galaxy catalogue \citep{Dalya2018, Dalya:2021ewn}, a 22-million object compilation from the Gravitational Wave Galaxy Catalogue (GWGC) \citep{GWGC}, 2MASS Photometric Redshift catalogue (2MPZ) \citep{2MPZ_2013}, 2 Micron All-Sky Survey Extended Source Catalogue (2MASS XSC) \citep{2MASS_XSC_2012}, HyperLEDA \citep{HyperLEDA_2014}, and WISExSCOS Photometric Redshift Catalogue galaxy catalogues \citep{WISExSCOSPZ}, and the and the 16th data release of the SloanDigital Sky Survey (SDSS) \citep{SDSS}. The GLADE+ catalogue reports galaxy redshifts along with their spectroscopic/photometric uncertainties, and already includes corrections for peculiar velocities from \cite{Mukherjee:2019qmm} in the reported uncertainties for nearby galaxies.

We use $K_{\mathrm{S}}$-band (denoted $K$-band) in Vega system absolute magnitudes as our fiducial tracer of stellar mass, assigning each galaxy a host‑probability weight. To model catalogue incompleteness, we adopt a Schechter luminosity function \citep{Schechter} with characteristic magnitude $M_{*,K} = -23.39 + 5 \log(h)$ with $h=H_0/100$, slope $\alpha_K = -1.09$ \citep{Kochanek:2000im} and truncate at $M_{K,\text{min}} = -27.0+ 5 \log(h)$ and $M_{K,\text{max}} = -19.0+ 5 \log(h)$. We apply $K$-corrections following \cite{Kochanek2001}.
Apparent magnitude thresholds $m_\text{th}$, needed to compute the limits of the out-of-catalogue integral (in Eq.~\ref{eq:out-cat} below), are obtained per HEALPix pixel as a median apparent magnitude in a given pixel, $K$-band threshold of 13.5 on average.

\section{Method}
\label{sec:method}

In our analysis we use the \gwcosmo pipeline presented in \cite{Gray:2019ksv,Gray:2021sew,Gray:2023wgj}. It requires precomputing a line-of-sight (LOS) redshift prior, a prior on GW signal in redshift and direction. It is given by
\begin{align}
    &p(z|\Omega_i, \Lambda, s, I) = \nonumber \\&= p(G|\Omega_i, \Lambda, s, I)\iint p(z, M, m|G, \Omega_i, \Lambda, s, I)dMdm \nonumber \\&+p(\bar{G}|\Omega_i, \Lambda, s, I)\iint p(z, M, m|\bar{G}, \Omega_i, \Lambda, s, I)dMdm
    \label{eq:los-z}
\end{align}
The first term denotes the in-catalogue part with the factor $p(G|\Omega_i, \Lambda, s, I)$ being the probability of the host galaxy being present in the galaxy catalogue given priors on the sky location $(\Omega_i)$, the cosmological and population hyperparameters of interest $(\Lambda)$, the presence of a GW source $(s)$ and additional
assumptions not explicitly expressed ($I$). The integral marginalizes over the possible host galaxies. The term inside the integral is the prior on $z$, $M$ and $m$ for host galaxies, informed by the galaxy catalogue, within the sky area covered by pixel $i$. The integral thus becomes a sum over the host galaxies,
\begin{align}
    &\iint p(z, M, m|G, \Omega_i, \Lambda, s, I)dMdm = \nonumber \\&=
    \frac{1}{p(s|G, \Omega_i, \Lambda, I) N_{\text{gal}}(\Omega_i)}\sum_i^{N_{\text{gal}}(\Omega_i)}p(z|\hat{z_i})p\left(s|z, M(z, \hat{m}_i, \Lambda), \Lambda, I\right) \text{,}
    \label{eq:in-cat}
\end{align}
The galaxies are treated as point sources modeled by a Gaussian and given weights which are some function of redshift and absolute magnitude, making the in-catalogue part essentially a weighted sum over the galaxies in the galaxy catalogue. 

The second term is the out-of-catalogue correction, with the factor $p(\bar{G}|\Omega_i, \Lambda, s, I)$ being the probability that the host galaxy is not present in the galaxy catalogue given the priors in $\Omega_i$, $\Lambda$, $s$ and $I$. The integral marginalizes over the possible host galaxies not present in the galaxy catalogue, and can be simplified to 
\begin{align}
    &\iint p(z, M, m|\bar{G}, \Omega_i, \Lambda, s, I) = \nonumber \\ 
    & = \frac{1}{p(s|\bar{G}, \Omega_i, \Lambda, I)p(\bar{G}|\Omega_i, \Lambda, I)} \nonumber \\    
    & \quad \times \left[ \int_{M(z, m_{\text{th}}(\Omega_i), \Lambda)}^{M_{\text{max}}(H_0)} p(z, M|\Lambda, I)p(s|z, M, \Lambda, I)dM
    \right]
    \label{eq:out-cat}
\end{align}
where $p(z, M|\Lambda, I)$ is the luminosity function of galaxies, usually given by the Schechter function and $p(s|z, M, \Lambda, I)$ is a weighting term. The limits of the integration are given by the dimmest and the brightest galaxy in the model. 

\begin{figure}
    \centering
    \includegraphics[width=0.8\linewidth]{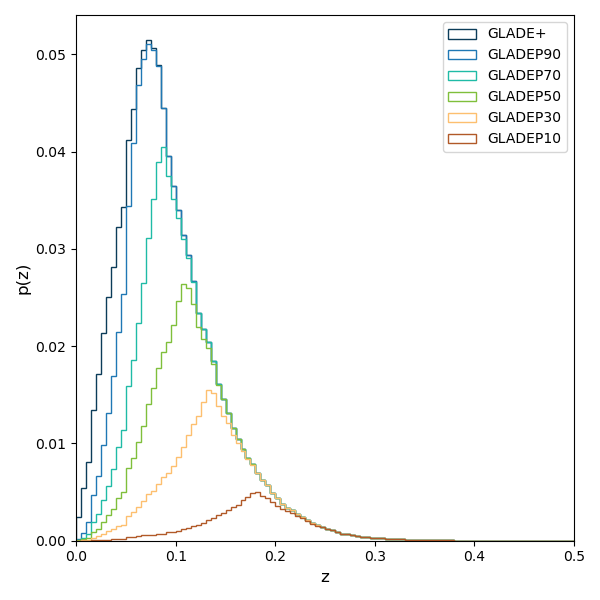}
    \caption{Normalized redshift distributions $p(z)$ for the full GLADE+ galaxy catalogue (blue) and the brightest percentiles (other curves). Each subset, labeled “GLADEPXX,” includes only the top XX\% brightest galaxies in the $K$-band. Focusing on increasingly brighter subsets shifts and extends the redshift distribution, effectively probing deeper ranges in $z$.}
    \label{fig:z_dist}
\end{figure}

To limit our analysis to bright galaxies, we choose a number of subsets of the GLADE+ catalogue containing only the brightest percentiles of galaxies. 
This sets the dimmest galaxy in the subset we use. We then apply its magnitude as the limit in integration over the out-of-catalogue part, assuming that the bright galaxies are the tracer of the mass in the universe. This makes the out-of-catalogue part smaller, effectively making the galaxy catalogue more complete in the high redshift ranges, as shown in Figure \ref{fig:z_dist}.

\section{Preservation of large-scale structure}
\label{sec:angular_power}

To assess whether selecting only bright galaxy subsets alters the large-scale structure traced by the galaxy catalogue, we compute the angular power spectrum of galaxy number density fluctuations in fixed redshift bins.

\subsection{Angular power spectrum estimation}
For a given redshift slice, we pixelize the sky using HEALPix and construct a galaxy overdensity field
\begin{align}
    \delta\left(\hat{n}\right) = \frac{N\left(\hat{n}\right) - \langle N\left(\hat{n}\right)\rangle}{\langle N\left(\hat{n}\right)\rangle}
\end{align}
where $N\left(\hat{n}\right)$ is the number of galaxies in a pixel and $\langle N\left(\hat{n}\right)\rangle$ is an estimate of the local mean galaxy density. The expectation map $\langle N\left(\hat{n}\right)\rangle$ is estimated from a low-resolution map that captures the survey selection function on the sky; this map is then upgraded to the same resolution as the HEALPix map and converted into the expected counts per pixel. Pixels with 
 $\langle N\rangle = 0$ define the angular mask. We apodize the mask before estimating the power spectrum.

The angular power spectrum $C_\ell$ is computed using the pseudo-$C_\ell$ estimator implemented in \texttt{NaMaster} \citep{namaster}, which correctly accounts for incomplete sky coverage and the resulting mode coupling induced by the survey mask. The spectra are binned in multipole space with bin width $\Delta \ell = 10$ and we present results in terms of $\ell(\ell +1)C_\ell/2\pi$.

The measured bandpowers include a shot-noise bias from discrete sampling. We estimate the mean shot-noise bandpower bias $N_\ell$ with a Monte Carlo procedure: for the fixed mask and selection function, we generate Poisson realisations of the counts map, compute the corresponding overdensity field, and run \texttt{NaMaster} to obtain pseudo-$C_\ell$ bandpowers. Averaging over these Poisson realisations yields an estimate of $N_l$, which is subtracted from the observed bandpowers to obtain an estimate of the signal bandpowers.

\subsection{Redshift uncertainties}
To propagate catalogue redshift uncertainties into the angular power spectrum, we use the per-galaxy redshift $z_i$ and uncertainty $\sigma_z$ provided in the catalogue. Rather than treating redshift bin membership as exact, we perform a Monte Carlo resampling in which each galaxy redshift $z'_i$ is now drawn from a Gausssian distribution $\mathcal{N}$ with the above values for mean and standard deviation,
\begin{align}
    z'_i \sim \mathcal{N}\left( z_i, \sigma_{z,i}\right),
\end{align}
after which galaxies are re-selected into the appropiate redshift bin and the angular power spectrum is recomputed following the same procedure as for the fiducial catalogue. Repeating this process yields an ensemble of angular power spectra, whose standard deviation at each multipole $\ell$ quantifies the uncertainty induced by redshift errors, denoted $\sigma_{\sigma_z}(\ell)$.

The total uncertainty on the angular power spectrum is computed as
\begin{align}
    \sigma_{\mathrm{tot}}(\ell) = \sqrt{\sigma_{\mathrm{Knox}}^2(\ell) + \sigma_{\sigma_z}^2(\ell)},
\end{align}
where $\sigma_{\mathrm{Knox}}$ is the standard Knox approximation \citep{Knox_1995},
\begin{align}
    \label{eq:knox}
    \sigma_{\mathrm{Knox}}(\ell) \simeq 
    \sqrt{\frac{2}{(2\ell+1)\,\Delta\ell\,f_{\mathrm{sky}}}}\,
    \left(C_\ell + N_\ell\right),
\end{align}
which accounts for cosmic variance and finite mode sampling on a partial sky, and $\sigma_{\sigma_z}$ captures the additional variance induced by galaxy redshift uncertainties. Since these two contributions arise from independent physical effects, they are added in quadrature.

In practice, we compute $\sigma_{\mathrm{Knox}}(\ell)$ using Eq. \ref{eq:knox} evaluated at the measured bandpowers, with $f_{\mathrm{sky}}$ determined from the apodized survey mask.

As a cross-check, we also estimate the full bandpower covariance using the Gaussian covariance formalism implemented in \texttt{NaMaster}, which explicitly accounts for mode coupling induced by the survey mask. We find excellent agreement between the diagonal elements of the \texttt{NaMaster} covariance and the Knox-based uncertainty estimate over the multipole range considered, validating the use of the Knox approximation for our analysis.

We note that our analysis focuses on multipoles $\ell \gtrsim 15$, such that the largest angular scales, where cosmic variance would normally dominate, are not probed. As a result, the total uncertainty does not exhibit the sharp increase typically seen at very low multipoles.

At higher multipoles, the uncertainties increase due to a combination of effects, including the growing impact of shot noise from the finite galaxy number density, particularly for the sparser magnitude-selected subsets, as well as limitations associated with the finite angular resolution of the catalogue and the HEALPix pixelization. Mask apodization and residual mode coupling further suppress small-scale power, effectively limiting the highest multipoles that can be robustly interpreted.

We find that redshift uncertainties introduce only a modest additional scatter in the angular power spectrum on the angular scales relevant for our analysis. In practice, the total uncertainty budget is dominated by the Knox term, which accounts for cosmic variance and the finite number of accessible modes on the partial sky, while the contribution from redshift uncertainties constitutes a subdominant correction.

\subsection{Choice of redshift bins}
We focus on two redshift bins, $0<z\leq 0.1$ and $0.1<z\leq 0.2$. These bins contain the majority of galaxies in the GLADE+ catalogue and dominate the statistical constraining power of the angular clustering measurements. For the brightest (and hence sparsest) subsets, the latter bin contains a larger fraction of the galaxies, while higher-redshift bins are increasingly affected by shot noise due to low number counts. We therefore restrict our analysis to these two bins, which capture the relevant large-scale structure signal while maintaining sufficient statistical robustness.

\subsection{Comparison}

\begin{figure}
    \centering
    \begin{subfigure}{\linewidth}
        \centering
        \includegraphics[width=0.8\linewidth]{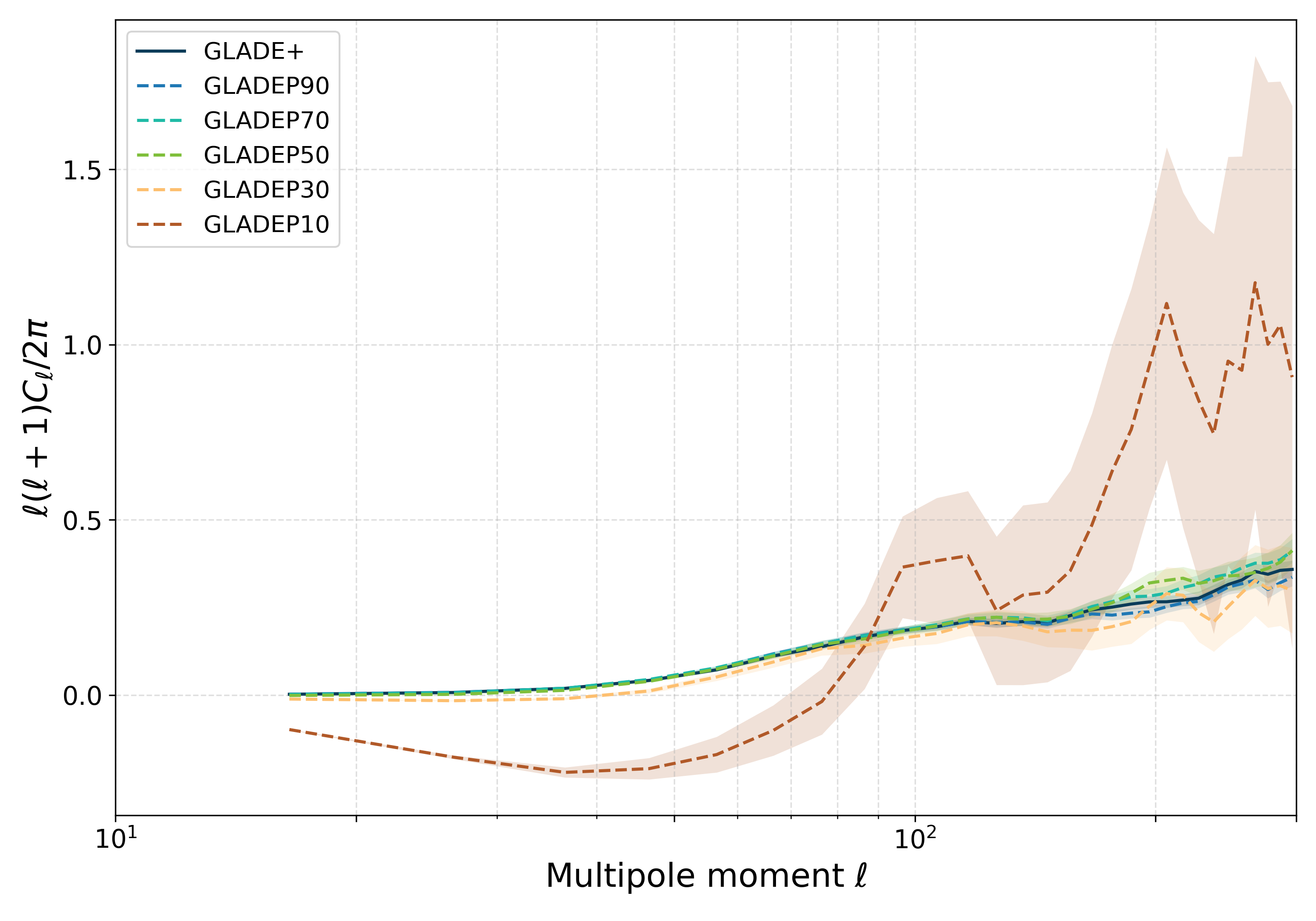}
        \caption{$0 < z \le 0.1$}
        \label{fig:LSS_zlow}
    \end{subfigure}

    \vspace{0.3cm}

    \begin{subfigure}{\linewidth}
        \centering
        \includegraphics[width=0.8\linewidth]{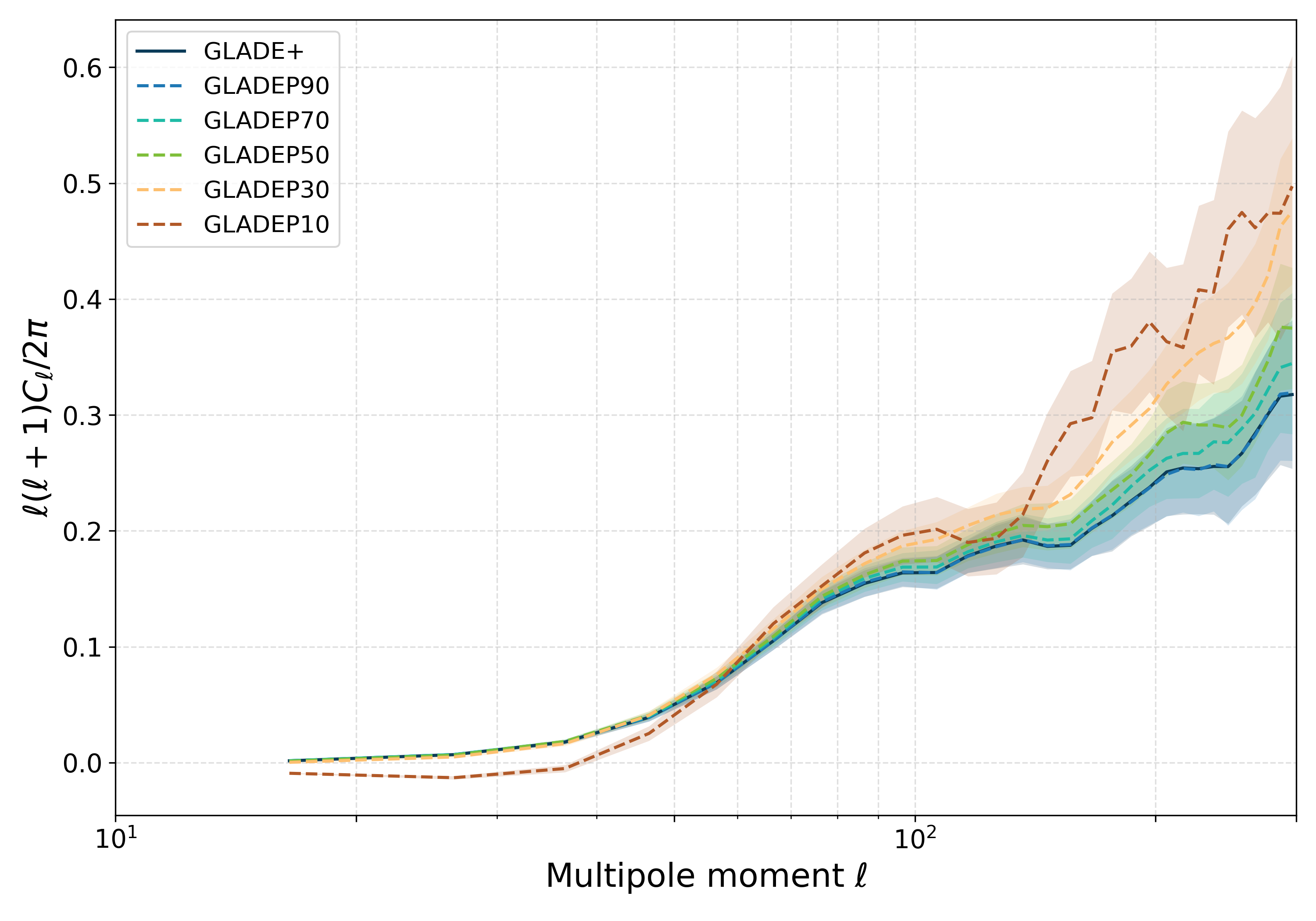}
        \caption{$0.1 < z \le 0.2$}
        \label{fig:LSS_zhigh}
    \end{subfigure}

    \caption{
    Angular power spectra of galaxy number density fluctuations for the full GLADE+ catalogue and brightness-selected subsets.
    The top panel shows the low-redshift bin $0 < z \le 0.1$, while the bottom panel shows $0.1 < z \le 0.2$, which contains the majority of galaxies for the stricter magnitude cuts.
    Shaded bands indicate the total uncertainty, combining cosmic variance and finite mode sampling (Knox approximation) with the additional variance induced by redshift uncertainties.
    Across both redshift bins, the amplitude and scale dependence of the angular power spectrum remain consistent between the full catalogue and bright subsets, until as low as $30\%$ of the brightest galaxies are chosen, demonstrating that magnitude cuts do not significantly alter the large-scale structure traced by the catalogue. For the extremely sparse subset with only $10\%$ of the brightest galaxies, the spectrum is dominated by shot noise, and any associated results are expected to be untrustworthy.}
    \label{fig:LSS}
\end{figure}

We directly compare the angular power spectra of the magnitude-selected subsets to that of the full galaxy catalogue within the same redshift bins. Crucially, the angular power spectra of the magnitude-selected subsets closely track that of the full catalogue over the multipole range considered. We find that even for very sparse subsets, corresponding to completeness levels as low as 30\%, the angular power spectra remain consistent with that of the full catalogue within the estimated uncertainties (Figure~\ref{fig:LSS}). In particular, both the overall amplitude and the scale dependence of the clustering signal are preserved, demonstrating that the underlying large-scale structure is conserved when restricting the catalogue to bright galaxies. Only for the most stringent and sparsest magnitude cuts do mild deviations appear, which can be partly attributed to increased shot noise from reduced galaxy number density. Overall, these results show that selecting bright subsets preserves the large-scale structure traced by the galaxy distribution. Since the power spectra for the GLADEP10 case (most stringent cut) show a different behaviour, it may not be advisable to use this subset for cosmology inference.

\section{Results}
\label{sec:results}

\subsection{Line-of-sight redshift prior}

\begin{figure}
    \centering
    \includegraphics[width=0.8\linewidth]{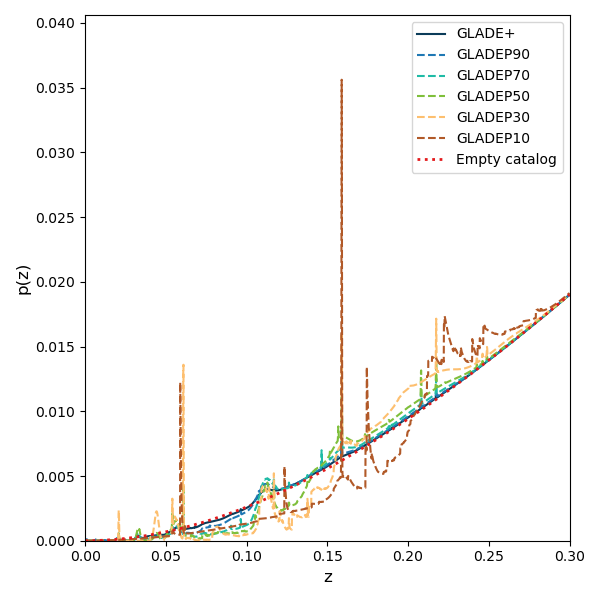}
    \caption{LOS redshift prior for GW170809, using the complete GLADE+ galaxy catalogue (blue), the brightest percentiles (dashed curves), and empty catalogue (red). Applying a brightness cut amplifies the prior at higher distances, extending the effective reach of the catalogue and partially mitigating incompleteness at greater redshifts.}
    \label{fig:zprior_event}
\end{figure}

Our analysis begins by constructing the LOS redshift prior, which is critical for the dark siren method. We decompose the prior into two distinct contributions: the in-catalogue component and the out-of-catalogue correction. For the in-catalogue term, we compute a weighted sum over the galaxies using luminosity weighting in the $K$-band, as described in the previous section. The weighting scheme assigns greater influence to brighter galaxies, based on the rationale that these objects are tracers of the underlying mass distribution. The integral in the out-of-catalogue correction is limited to the maximum magnitude of the catalogue subset, effectively making the catalogue more complete.

As shown in Figure \ref{fig:zprior_event} for GW170809, the application of a brightness cut significantly modifies the LOS redshift distribution. Specifically, the bright galaxy subsets yield an amplified redshift prior at higher distances, effectively extending the reach of the catalogue, somewhat mitigating the incompleteness issues that arise at deeper redshifts.

\subsection{Constraints on $H_0$}

\begin{figure}
    \centering
    \includegraphics[width=0.8\linewidth]{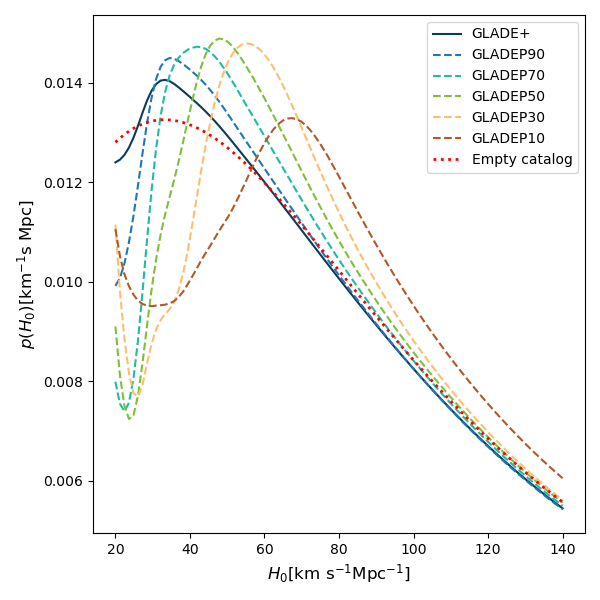}
    \caption{Posterior distributions of the Hubble constant, for GW170809, for various brightness selections from the galaxy catalogue.}
    \label{fig:H0_event}
\end{figure}

The enhanced LOS prior directly impacts our estimation of the Hubble constant. With a more complete and deeper effective galaxy catalogue, our dark siren analysis benefits from improved statistical power. The weighted contribution of bright galaxies leads to tighter constraints on the Hubble constant, as demonstrated in Figure \ref{fig:H0_event} by the shift in the posterior distributions compared with those obtained using the complete catalogue. Our results indicate that the brightness selection not only increases the effective depth of the galaxy sample but also reduces uncertainties in the inferred cosmological parameters. 

\begin{figure}
    \centering
    \includegraphics[width=0.8\linewidth]{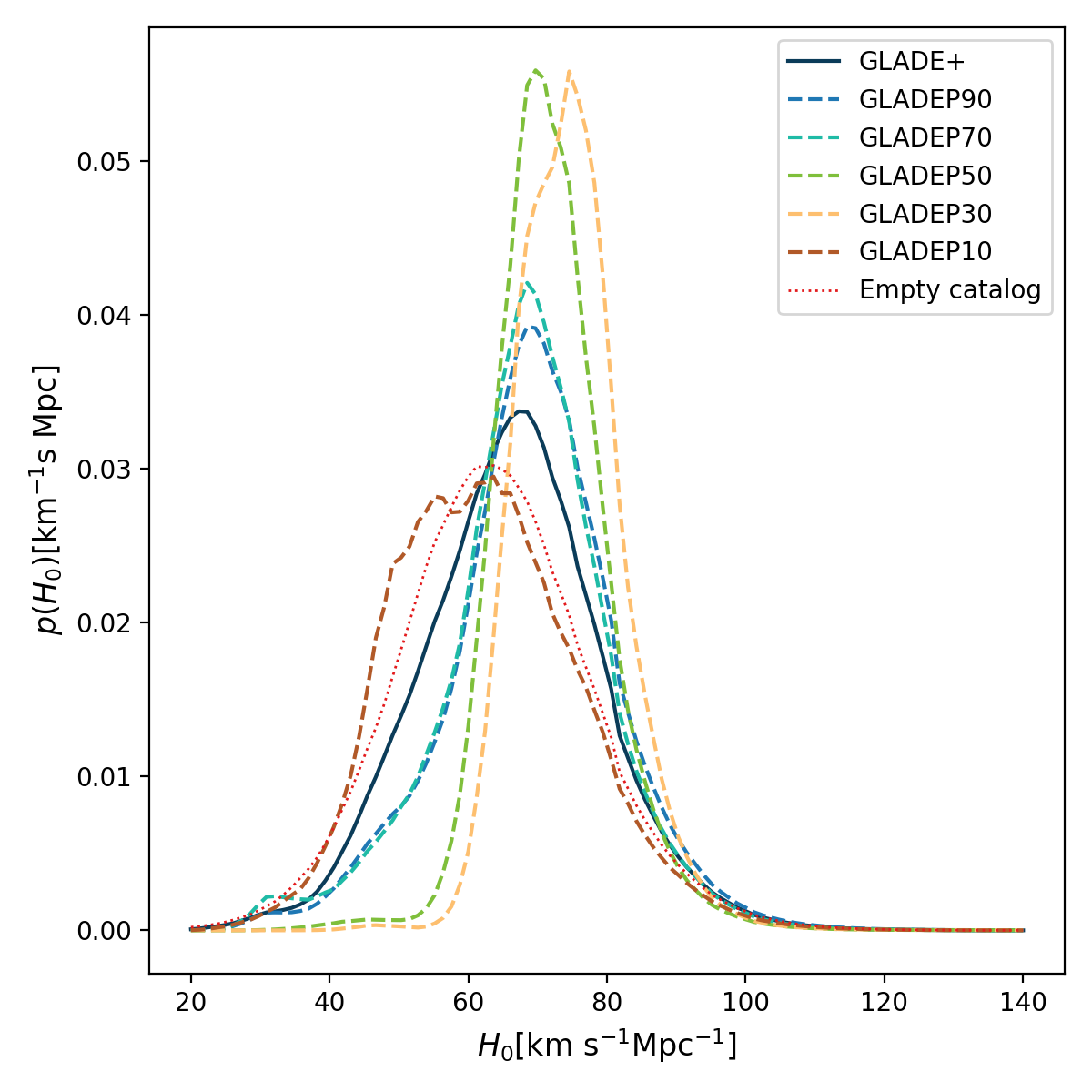}
    \caption{Posterior distributions of the Hubble constant for various brightness cuts. Moderate brightness cuts yield tighter $H_0$ constraints by effectively extending the catalogue’s depth, while overly restrictive cuts (e.g., using only the top $10\%$ of brightest galaxies) can lead to an underpopulated sample, diminishing the statistical power of the inference.}
    \label{fig:H0_all}
\end{figure}

The combined analysis for the selected GW events yields similar results, with the use of brightest subsets leading to much tighter constraints and a slightly higher median value for $H_0$ as compared to the complete GLADE+, and empty catalogue estimates (see Figure \ref{fig:H0_all}). These results are also tabulated in Table~\ref{tab:H0_all}. In the most optimistic case (GLADEP20), we see an improvement by a factor of $80\%$ over the full GLADE+ result.

However, there is a practical limit to how restrictive the brightness cut can be. While focusing on an extremely bright subset (e.g., the top 10\% of brightest galaxies, GLADEP10) further increases the average completeness at higher redshifts, it can leave the catalogue effectively empty at certain redshift ranges due to sparse galaxy counts. This occurs because there are too few galaxies that meet the stringent brightness threshold, leading to reduced statistical power in the final cosmological inference. This result is also in agreement with our expectations from Section~\ref{sec:angular_power}, where we have seen in the computed angular power spectra that the GLADEP10 subset does not preserve the underlying large-scale structure. Thus, a balance must be struck between selecting sufficiently bright galaxies to probe deeper redshifts and retaining enough galaxies to avoid an underconstrained analysis.

\begin{table}
    \centering
    \caption{Maximum \textit{a posteriori} values with $68\%$ credible intervals of the $H_0$ posterior distributions for different magnitude-selected percentiles of the GLADE+ galaxy catalogue. Results are shown for both luminosity-weighted and uniformly weighted galaxy priors, together with the corresponding maximum magnitude limits.}
    \begin{tabular}{ c c c c }
    \hline
        Catalogue & $M_{K, max}$ & $H_0$ & $H_0^{\mathrm{uniform}}$ \\
          &              & $[\mathrm{km~s^{-1}Mpc^{-1}}]$ & $[\mathrm{km~s^{-1}Mpc^{-1}}]$ \\ \hline
        GLADE+ & -19.00 & $67.87^{+8.97}_{-10.29}$ & $63.85^{+9.49}_{-11.00}$ \\ \hline
        GLADEP90 & -23.07 & $68.94^{+9.24}_{-7.55}$ & $69.00^{+9.69}_{-7.79}$ \\ \hline
        GLADEP80 & -23.62 & $68.93^{+9.25}_{-7.57}$ & $68.97^{+10.22}_{-7.75}$ \\ \hline
        GLADEP70 & -23.94 & $68.63^{+8.61}_{-7.42}$ & $68.45^{+8.52}_{-9.44}$ \\ \hline
        GLADEP60 & -24.19 & $68.85^{+7.72}_{-6.43}$ & $68.68^{+8.02}_{-7.47}$ \\ \hline
        GLADEP50 & -24.14  & $70.05^{+6.12}_{-5.20}$ & $70.21^{+5.93}_{-5.36}$ \\ \hline
        GLADEP40 & -24.63 & $69.94^{+7.46}_{-3.88}$ & $70.50^{+6.47}_{-4.65}$ \\ \hline
        GLADEP30 & -24.87 & $74.70^{+4.69}_{-6.58}$ & $74.41^{+3.88}_{-7.14}$ \\ \hline
        GLADEP20 & -25.15 & $78.28^{+5.53}_{-4.95}$ & $78.21^{+5.59}_{-4.88}$ \\ \hline
        GLADEP10 & -25.53 & $63.46^{+6.39}_{-14.37}$ & $63.46^{+6.24}_{-14.90}$ \\ \hline
        Empty catalogue & - & $63.63^{+6.07}_{-15.07}$ & $63.63^{+6.07}_{-15.07}$ \\ \hline
    \end{tabular}
    \label{tab:H0_all}
\end{table}

\subsection{Uniform galaxy weighting as a robustness test}
As an additional systematic check, we repeat the dark siren analysis using uniform galaxy weights, instead of luminosity-based weighting in the $K$-band. In this case, all galaxies within a given catalogue subset contribute equally to the line-of-sight redshift prior, removing any explicit dependence on galaxy luminosity or stellar-mass proxies.

\begin{figure}
    \centering
    \includegraphics[width=0.8\linewidth]{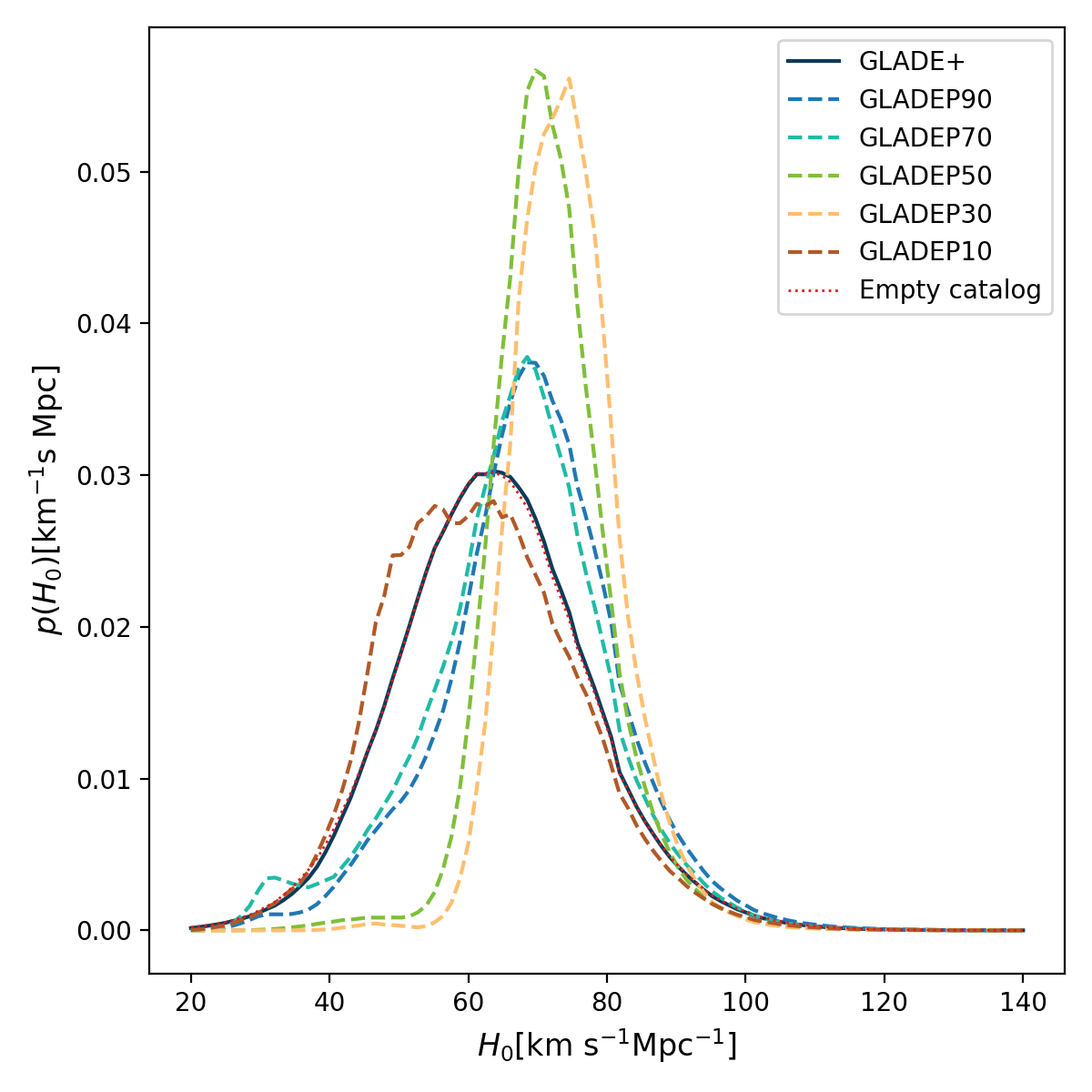}
    \caption{
Posterior distributions of the Hubble constant obtained using \emph{uniform} galaxy weighting for the full GLADE+ catalogue and brightness-selected subsets.
In contrast to luminosity weighting, the uniformly weighted full-catalogue result closely resembles the empty-catalogue posterior, indicating that numerous faint galaxies contribute little constraining power when treated equally.
Applying brightness cuts recovers tighter constraints, as the analysis becomes dominated by fewer, more informative galaxies.
This behaviour mirrors the effect observed with luminosity weighting and demonstrates that both luminosity weighting and magnitude selection act to emphasize the same large-scale structure information relevant for dark siren cosmology.}
    \label{fig:uniform_weighting}
\end{figure}

We find that the resulting $H_0$ posterior distributions are consistent with those obtained using luminosity weighting, across all brightness-selected subsets (Figure~\ref{fig:uniform_weighting}). While uniform weighting generally leads to slightly broader posteriors due to the loss of information on the relative importance of galaxies, the qualitative behavior remains unchanged: moderate brightness cuts still improve the constraining power relative to the full catalogue, and extreme cuts lead to diminished performance due to sparse sampling.

We note an important qualitative difference between the uniformly weighted and luminosity-weighted analyses for the full galaxy catalogue. When uniform weights are adopted, the resulting $H_0$ posterior closely resembles that obtained using an empty catalogue, indicating that the full catalogue, dominated by numerous faint and weakly informative galaxies, adds limited constraining power in this case.

In contrast, luminosity weighting naturally enhances the contribution of brighter, typically closer galaxies, leading to a posterior that is narrower. Notably, this behavior mirrors the effect observed when explicitly selecting bright galaxy subsets. This demonstrates that luminosity weighting and magnitude cuts act in a qualitatively similar manner, by preferentially emphasizing the galaxies that most effectively trace the local large-scale structure relevant for dark siren measurements.

This test also corroborates our interpretation that the small variations observed in the angular power spectrum and in the inferred $H_0$ constraints are driven primarily by the underlying large-scale structure traced by the catalogue, rather than by the specific choice of galaxy weighting scheme.

\subsection{Cost-benefit trade-off}
\label{sec:mdc}

While the use of a bright subset enhances the completeness of the catalogue at high redshifts, it also introduces a trade-off. If the brightness cut is too restrictive, the exclusion of fainter galaxies may lead to an underrepresentation of the overall matter distribution, potentially biasing the results. One needs to carefully balance these considerations by comparing different brightness thresholds. A moderate brightness cut will maximize the benefit in terms of depth without incurring significant bias. The fact that the angular power spectrum remains unaltered to completeness levels as low as 30\%, already gives us confidence that the result is fairly unbiased until this level.

An astrophysics-motivated set of end-to-end simulations or a GW ``mock data challenge''  with comparable detector sensitivities and galaxy catalogue parameters will further establish the robustness of these results or else identify caveats in certain regimes. Challenges are matching parameters of ideal simulated catalogues to realistic ones, in particular, composite catalogues such as GLADE+. A study with the simulated ``BUZZARD'' galaxy catalogue from the Dark Energy Survey Collaboration~\cite{DES:2019jmj,DES:2021bwg} is already being planned. Tests performed in the process will help refine the methodology and ensure that the improved constraints on the Hubble constant are robust against additional systematic uncertainties.

\section{Conclusions}
\label{sec:conclusions}

Our study demonstrates that choosing only bright galaxies to provide redshift support for a dark standard siren measurement of $H_0$ can improve the measurement precision by reducing the relative weight of the out-of-galaxy catalogue contribution. A thorough study is needed to better quantify an optimal fraction of galaxies to choose and further assess the systematic effects of the method.

Bright galaxies provide complementary information to other tracers at high redshifts, such as galaxy clusters~\citep[see][]{Beirnaert:2025wcx} or HI intensity mapping. Using such alternative tracers are not only useful but \textit{necessary} in conjunction with GW data with enhanced LVK sensitivities, where numerous BBHs are expected at distances corresponding to $z\sim2$ and future observatories such as the Einstein Telescope~\citep{Abac:2025saz} and the Cosmic Explorer~\citep{Evans:2021gyd} as well as the space-based LISA~\citep{LISA:2024hlh}, which push the detection horizon well-beyond $z\gtrsim3$.

\section*{Acknowledgements}

We would like to thank Gergely Dálya for a careful review of the manuscript and Freija Beirnaert, Maciej Bilicki, Rachel Gray, Simone Mastrogiovanni, Nicola Tamanini, and other members of the LVK CBC Cosmology working group for useful thoughts and comments throughout the course of this study. Most importantly, we are grateful to the GWCats consortium for a workshop during which this project was conceived and several meetings in which this work and related aspects were presented and discussed, in particular, with Felipe Andrade Oliveira and Marcelle Soares-Santos.

Our research is supported by Ghent University Special Research Funds (BOF) project BOF/STA/202009/040, the inter-university iBOF project BOF20/IBF/124, and the Fonds Wetenschappelĳk Onderzoek (FWO) research project G0A5E24N. We also acknowledge support to the FWO International Research Infrastructure (IRI) grant \textit{``Essential Technologies for the Einstein Telescope''} (I002123N) for funding related to LVK membership and travel.

This material is based upon work supported by NSF's LIGO Laboratory which is a major facility fully funded by the National Science Foundation in the U.S.A.~and Virgo supported by the European Gravitational Observatory (EGO) and its member states. The authors are grateful for computational resources provided by the LIGO Laboratory and supported by National Science Foundation Grants PHY-0757058 and PHY-0823459.

This work makes use of \gwcosmo which is available at \url{https://git.ligo.org/lscsoft/gwcosmo}.

\section*{Data Availability}

The GLADE+ catalogue is available at \hyperlink{https://glade.elte.hu/}{https://glade.elte.hu/}.
All of the GW events we have used in this analysis are available at \hyperlink{https://www.gw-openscience.org/}{https://www.gw-openscience.org/}.

\bibliographystyle{mnras}
\bibliography{bibliography}

\bsp	
\label{lastpage}
\end{document}